\documentclass{emulateapj}
\usepackage{natbib}

\newcommand \msun {\mathrm{M}_\odot}
\newcommand \degree {\ensuremath{^\circ}}

\begin{document}

\title{Direct Detection of the Brown Dwarf GJ 802B with
Adaptive Optics Masking Interferometry}
\author{James P. Lloyd\altaffilmark{1}}
\author{Frantz Martinache\altaffilmark{1}}
\author{Michael J. Ireland\altaffilmark{2}}
\author{John D. Monnier\altaffilmark{3}}
\author{Steven H. Pravdo\altaffilmark{4}}
\author{Stuart B. Shaklan\altaffilmark{4}}
\author{Peter G. Tuthill\altaffilmark{5}}

\altaffiltext{1}{Department of Astronomy, Cornell University, Ithaca
NY}
\altaffiltext{2}{Division of Geological and Planetary Sciences,
California Institute of Technology, Pasadena CA}
\altaffiltext{3}{University of Michigan Astronomy Department, 941 Dennison Building, Ann Arbor, MI}
\altaffiltext{4}{Jet Propulsion Laboratory, Pasadena, CA}
\altaffiltext{3}{School of Physics, University of Sydney, Sydney NSW
Australia}
\date{}

\begin{abstract}

We have used the Palomar 200" Adaptive Optics (AO) system to directly detect
the astrometric brown dwarf GJ 802B reported by Pravdo et al. 2005.
This observation is achieved with a novel combination of aperture
masking interferometry and AO.  The dynamical masses are
0.175$\pm$0.021 M$_\odot$ and 0.064$\pm$0.032 M$_\odot$ for the primary and
secondary respectively.  The inferred absolute H band magnitude of GJ
802B is M$_H$=12.8 resulting in a model-dependent T$_\mathrm{eff}$ of
1850 $\pm$ 50K and mass range of 0.057--0.074 M$_\odot$.   

\end{abstract}


\section{Introduction}

Binary stars provide a unique laboratory for the study of the
physical properties of individual objects, and important constraints
on star formation and evolution.  Although there are now a large
number of objects known beyond the substellar limit, there are few in
systems amenable to extraction of dynamical measurements of physical
parameters~\citep{2006astro.ph..2122B}.  This is a particularly acute
issue in the case of substellar objects due to the degeneracy between
age, mass and luminosity.

We have selected targets from the STEPS survey
\citep{1996ApJ...465..264P} that show astrometric evidence of a low
mass companion for imaging followup with Adaptive Optics (AO).  In
order to improve the sensitivity to companions we have implemented a
novel aperture masking interferometry technique in concert with the
atmospheric turbulence correction afforded by AO.  This technique has
succeeded in directly detecting the low mass companion to the M5.5
dwarf GJ 802.

\section{Observations}

\subsection{Aperture Masking Interferometry}
\label{sec:NRM}

While the large gains of AO for high contrast imaging are widely
recognised (e.g \citet{2003ApJ...594..538M, 2005AJ....130.1212C,
2005Natur.433..286C}), nearly all AO searches for substellar
companions have focussed on achieving very high dynamic range ($>
10^4$) at moderate separations.  Several issues drive AO imaging to
this parameter space.  The interaction of the actuator count of AO
systems and the limitations of coronagraphy
\citep{Sivaramakrishnan01,Lloyd01} lead to both practical and
fundamental limitations for high contrast imaging.  Present and
currently planned \citep{2004SPIE.5490..359M} AO systems are focussed
on achieving high contrast at radii of more than 4 $\lambda/D$.
At closer separations, diffracted light is difficult to suppress with
a coronagraph, and most importantly there is a large noise floor due
to the presence of fluctuating speckles in the image
\citep{1999PASP..111..587R, 2006ApJ...637..541F,
2006dies.conf..581S}.   Finally,  it has proven to be remarkably
difficult in practice to precisely calibrate the AO PSF.  A variety
of differential imaging approches have been proposed to circumvent
the problem of AO PSF calibration \citep{2000PASP..112...91M,
2004ApJ...615L..61M, 2005ASPC..343...75B, 2006ApJ...641..556M,
2002ApJ...578..543S}.  These approaches rely on exploiting a
differential signal in wavelength, polarization or sky rotation to
improve the source extraction, but do not fundamentally address the
issue of calibration of AO data.

\begin{figure}[htbp]
\begin{center}
\includegraphics[width=2.8in]{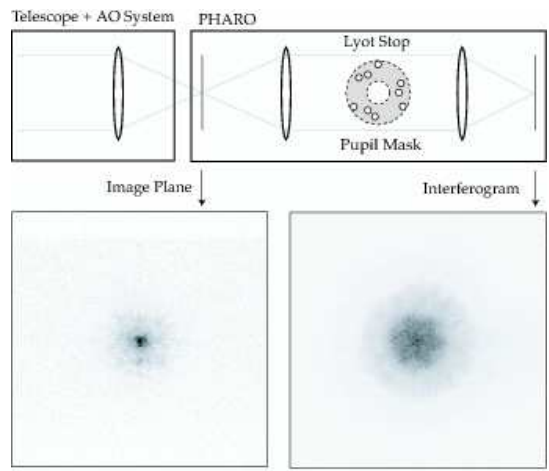}
\caption{Functional diagram of the aperture masking experiment.
Although in practice the optical system is complex and involves many
reflective optical elements, only the essential imaging properties
are represented with lenses.  After the telescope and AO system, the
telescope pupil is reimaged at the PHARO Lyot stop.  Under most
common circumstances, the Lyot stop serves only as a cold baffle.  In
a coronagraphy mode, an undersized stop is used to block the light
diffracted by the pupil edges.  For the aperture masking
interferometry mode,  a pupil mask is placed in the Lyot stop to form
an interferogram recorded at the focal plane.  The images shown are
the full pupil AO images and 9-hole interferograms obtained for GJ
802 in  2004 September.}
\label{fig:masking}
\end{center}
\end{figure}

In light of these considerations, we have undertaken a novel
experiment to achieve precision calibration of AO data, by marrying
the sensitivity of AO observations with the precision calibration
afforded by interferometry.  The heritage of Non-Redundant Masking
interferometry \citep{Tuthill00, Readhead88,Nakajima89} can be
combined with the wavefront stabilization of adaptive optics
\citep{Tuthill06}.  In practice, the optical implementation of this
capability is relatively simple (see Figure~\ref{fig:masking}).  We
have used the Palomar 200" telescope with the PALAO adaptive optics
system \citep{2000SPIE.4007...31T} and the PHARO infrared camera
\citep{2001PASP..113..105H}.  PHARO was designed with coronagraphic
capability in mind, so incorporates a ten position Lyot wheel in the
collimated beam at the internal position of the re-imaged telescope
pupil.   This wheel holds a variety of pupil stops to enable the
interchange of Lyot stops with various undersizings
\citep{2000SPIE.4007..899O}.  We have installed 
non-redundant masks in the spare openings of the PHARO Lyot wheel.
The 9-hole mask used in this work is
optimized for broad band, faint targets with 50 cm diameter
subapertures and 4.15 m longest baseline.  The 9-hole mask transmits
approximately 15\% of the total light incident on the telescope pupil.

\begin{figure}[htbp]
\begin{center}
\includegraphics[width=2.8in]{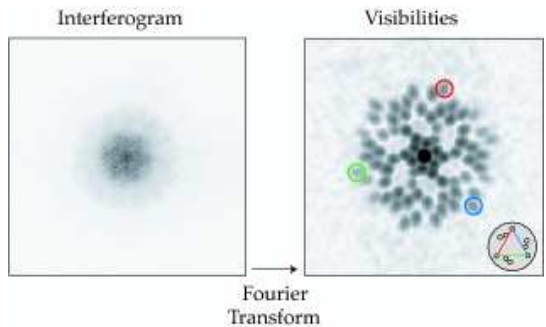}
\caption{Extraction of closure phases from optical interferograms.
Each ``splodge'' in the Fourier plane corresponds to the fringe
formed by a pair of holes in the pupil mask.  The splodges are
point-symmetric since the interferogram is real (or equivalently each
splodge appears for both the baseline formed by a pair of holes in
both order A-B and B-A).  Since the splodges map uniquely to a
baseline, closure phase relations can be constructed that reject any
residual phase errors, and therefore speckle noise.  One such closure
phase triangle (out of 84 possible for a 9 hole mask) is shown in the
lower right with the visibility splodges corresponding to each
baseline circled.}
\label{fig:closurephase}
\end{center}
\end{figure}

The advantages of this approach are severalfold.  By preserving
non-redundancy, a given baseline in the pupil formed by any pair of
subapertures translates uniquely to a single spatial frequency in the
detector plane.  The fringe observables are extracted from the
Fourier transform of the interferogram (see
Figure~\ref{fig:closurephase}).  Each observable is a fringe complex
visibility.  The preservation of the non-redundancy relation ensures
that the extracted fringes can be used to form closure phases
\citep{1958MNRAS.118..276J,Cornwell89}.  The compelling advantage of
the use of closure phase data is the rejection of any residual
pupil-plane phase errors, which are the source of both AO PSF
calibration difficulties and speckle noise.  The non-redundant
masking technique therefore rejects the phase noise associated with
both the instantaneous and time-averaged AO system performance.
Images can be reconstructed using self-calibration techniques.
Although the dynamic range achieved here is modest by comparison with
conventional AO it is uniquely close to the central
star, within a few $\lambda/D$, which is an area not accessible to
coronagraphs.  The use of closure relations in radio interferometry
has enabled imaging with dynamic range exceeding $10^5$ (e.g.
\citet{2003ApJ...593..169H}).

\subsection{GJ 802 observations}

GJ 802 was observed at the Palomar 200" telescope with conventional 
AO imaging and 9 hole aperture
masking interferometry on 2004 September 2 UT, in good seeing.
Uncompensated images earlier in the night showed 0\farcs6 FWHM seeing
at H band (0\farcs75 in V band).  The conventional AO imaging placed an upper limit of a
contrast ratio of 0.05 for any companion at $\sim$100 mas
\citep{Pravdo05}.  Imaging observations have also been attempted with
Keck/LGS (C. Gelino, pers. comm) and HST (GO-10517) but have not
detected the companion.

Interferograms were recorded using the Fowler sampling mode of PHARO
on a 256$\times$256 subarray.  PHARO provides a mode whereby all
reads of a Fowler sampling sequence can be saved.  We use this mode
and the minimum exposure time to save a data cube of sixteen
sequential non-destructive reads of the detector without reset.  This
provides fifteen pairwise 431 ms exposures in a total exposure time
of 6465 ms, thus very efficiently recording a large number of short
exposures, so long as the detector does not saturate in the total
exposure time.   For these observations, we recorded 115 image cubes
yielding 1725 431 ms exposures on source.  The large number of frames
allows good estimation of the errors.  The observations show
0.6\degree~RMS closure phase scatter.

Calibration of the interferograms is achieved by observing an unresolved source to
measure the system visibilities.   
It is usually 
considered necessary to choose a calibrator that is as similar as possible to
the target star in both wavefront sensor (approximately R band) and
science camera (H band) brightness, at similar airmass.  We select
calibrator stars by searching the USNO CCD Astrograph Catalog (UCAC2)
\citep{2004AJ....127.3043Z} and 2MASS catalog for stars nearby in the
sky with similar properties.  The UCAC2 catalog bandpass is between V
and R bands, and for practical purposes has proven to be a similar
magnitude scale to the PALAO wavefront sensor.  
For these
observations 2MASS 20494024+4526398  (2UCAC 47204238; UC=14.72 mag;
H=7.73 mag) was selected as a calibrator star.

\subsection{Data Analysis}

\begin{figure}[htbp]
\begin{center}
\includegraphics[width=2.8in]{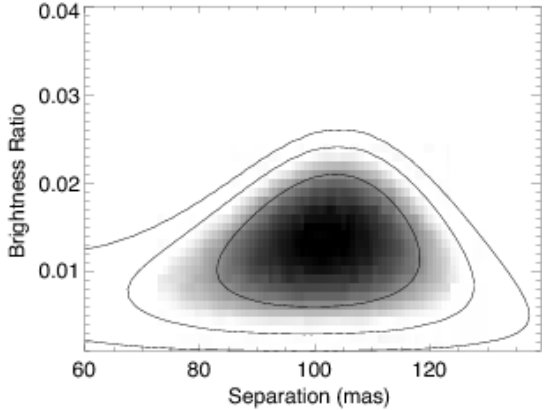}
\caption{Likelihood function cross-section for GJ 802 binary model
fit.  The contours represent 90, 99, and 99.9\% confidence levels.}
\label{fig:gj802-likelihood}
\end{center}
\end{figure}

The data was dark subtracted, flatfielded, and analysed with a custom
software pipeline written in IDL.  The pipeline outputs a bispectrum
in OIFITS format \citep{Pauls05}.   A binary model is fit to the
bispectrum with a reduced $\chi^2$ method.  
In practice we have found that the visibility amplitude
calibration is poor, and superior results are achieved with a fit to
the closure phase alone.  Presumably this is because the visibility
amplitude calibration is susceptible to the same fluctuations in
seeing and AO performance between source and calibrator that plague
conventional imaging with AO.   As discussed in section~\ref{sec:NRM}
the closure phase rejects the residual phase errors and is therefore
expected to be robust.  Although in principle the closure phase is self-calibrating, 
there are systematic non-zero closure phase errors of a few degrees.  Therefore it remains necessary to calibrate the 
non-zero closure phases.  The source of 
these non-zero closure phases is not entirely understood, but the level is consistent with the
expected telescope and AO system residual wavefront errors and detector flatfielding errors. 
Once the visibility amplitude 
is rejected from the analysis, it is also possible to include additional 
calibrators observed throughout the night.  These additional calibrators 
usually improve the estimation of errors, and provide a robustness against 
possibility that the calibrator itself is an unknown binary.  We include 
observations of HD 4915 (G0V, V=6.76, H= 5.416) 
HD 28005 (G0V, V= 6.71, H=5.506) as additional calibrators in this analysis.

Likelihood contours for the
binary model parameters are shown in
Figure~\ref{fig:gj802-likelihood}.  
The derived binary model is separation 102 $\pm$ 7 mas at position
angle 36.1 $\pm$ 4.5 degrees.  
The  H band contrast ratio is 74.4 +/- 18.5 ($\Delta$H = 4.68 $\pm$ 0.28 mag).

\begin{deluxetable}{ll}
\tablewidth{0pt}
\tablecaption{Orbital Elements\label{tab:orbit}
}
\tablehead{
\colhead{Quantity}       & \colhead{Value} }
\startdata
Absolute Parallax	       & 64.5 $\pm$ 2 mas\\
Proper Motion               & 1933 $\pm$ 1 mas\\
Position Angle	             & 26.6$\degree \pm 0.5\degree$ \\
Period                           & 3.13 $\pm$ 0.04 y\\
Total Mass                    & 0.24 $\pm$ 0.05 $\msun$ \\
Semi-major Axis          & 1.32 $\pm$ 0.09 AU\\
Eccentricity	            & 0.60 $\pm$ 0.29\\
Inclination                    & $82\degree \pm 2\degree$\\
Long. Of Asc. Node     & $18.5\degree \pm 4.5\degree$\\
Arg. Of Periastron       & $222.5\degree \pm 25.5\degree$\\
Epoch (y)	                  & 2000.51 $\pm$ 0.19\\
Primary Mass               & 0.175 $\pm$ 0.021 $\msun$\\
Secondary Mass           & 0.064 $\pm$ 0.032 $\msun$\\
\enddata
\end{deluxetable}
 
\section{Orbital Parameters}
 
\begin{figure}[htbp]
\begin{center}
\includegraphics[width=2.8in]{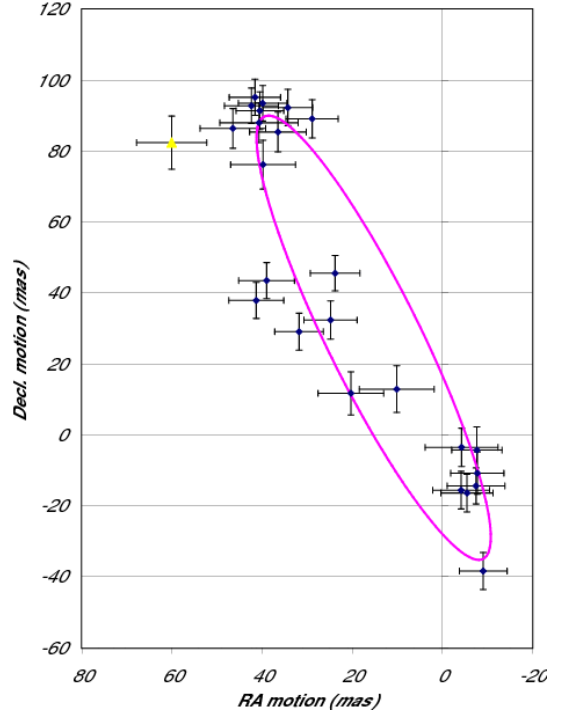}
\caption{Orbit of GJ 802B.  STEPS astrometric observations are
diamonds.  The
uncertainties shown are the photocenter astrometry uncertainties multiplied by the
ratio ($\sim$4) of the
displayed Keplerian orbit (curve) to the photometric orbit. The
resolved AO observation  is the triangle.
}
\label{fig:gj802-orbit}
\end{center}
\end{figure}

The measured position of GJ 802B, shown in
Figure~\ref{fig:gj802-orbit}, is very near the expected position from
\citet{Pravdo05} although this prior information is not involved in
the fit to the closure phases used to extract the astrometry.
Updated orbital elements combining this resolved observation and
additional astrometric observations since \citet{Pravdo05} are shown
in Table~\ref{tab:orbit}.  This orbit is consistent with the pure
photocenter astrometry orbit derived in \citet{Pravdo05}.

\section{Discussion}

\begin{figure}[htbp]
\begin{center}
\includegraphics[width=2.8in]{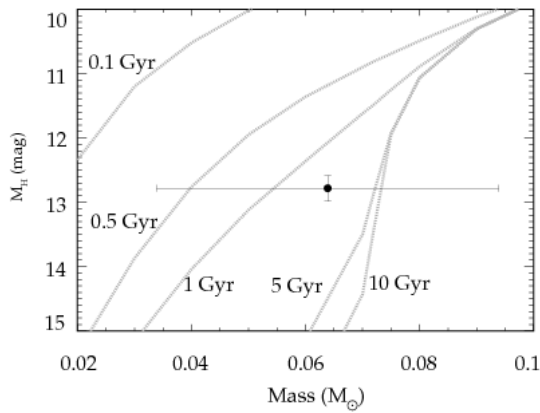}
\caption{Observed mass and luminosity of GJ 802B compared to
theoretical isochrones from \citep{Baraffe03}.  The mass is the
dynamical mass on the basis of the orbital fit shown in
Table~\ref{tab:orbit} and Figure~\ref{fig:gj802-orbit}.  Uncertainty
in the mass is large due to the poorly constrained orbital
eccentricity.  Future observations should constrain the eccentricity
and therefore the mass precisely.}
\label{fig:baraffe}
\end{center}
\end{figure}

The luminosity of GJ 802B can be determined precisely by a
differential measurement from GJ 802A.  The 2MASS catalog
\citep{2003tmc..book.....C} records the H band brightness of GJ 802
as 9.058$\pm$ 0.019 mag.   Adopting the parallax determined by
\citet{Pravdo05} of 64$\pm$2 mas, the absolute magnitude of GJ 802 is
M$_H$ = 8.11 $\pm$ 0.07 mag.  Using the  H band contrast ratio of
$\Delta$H = 4.68 $\pm$ 0.28 mag, we determine the absolute magnitude of GJ 802B to
be M$_H$ = 12.79$\pm$ 0.3.  Comparison with models of
\citet{Baraffe03} admits a large range of possible masses depending
on age (see Figure~\ref{fig:baraffe}), $0.057$--$0.074 \msun$, with
Teff = 1850$\pm$50K for models of ages $1$--$10$ Gyr.  With an age
estimate of $<$ 6 Gyr based on activity (Pravdo et al. 2005), the
Baraffe models indicate a mass $\sim 0.07~\msun$.  

For models with 
age $>$ 5 Gyr, the mass range consistent with this luminosity is remarkably 
narrow, $0.072$ -- $0.074$. This
model-dependent mass range is narrower than the present dynamical mass
determinations based on the STEPS orbit alone \citep{Pravdo05} or
this work.   Although this mass is consistent with the
orbital solution based on the STEPS and AO masking result, the high mass required by
the models if GJ 802 is old demands an unusually high eccentricity ($e>$0.8) for the astrometric 
orbital solution.   It is therefore tempting to conclude that the
GJ 802 is young ($<$ 1 Gyr), which would admit a lower eccentricity for the 
orbital solution.

A sample of field objects selected with a luminosity near the substellar limit would
be dominated by stellar objects since the cooling time of low mass stars dramatically
exceeds that of brown dwarfs.   Further, since the cooling time increases with mass, the 
distribution of field objects below the substellar limit contains many more old massive brown dwarfs
than young low mass objects. 
The conclusion that GJ 802 is young would be remarkably puzzling 
since then the only three resolved  binaries with dynamical masses below 0.08 $\msun$
(GJ 802B; GL 569 Bab \citep{2004ApJ...615..958Z,2006ApJ...644.1183S}; 2MASSW J0746425+2000321AB \citep{2004A&A...423..341B}) are 
inferred to be young brown dwarfs, despite the fact that young low mass objects are less likely to be found than old high mass ones.
These conclusions are suggestive that the models are 
under-predicting the luminosity of substellar objects. 

Ultimately,
further observatieoons will provide model-constraining dynamical
measurements of the masses of both the primary and secondary.  If GJ
802 is an old star, then GJ 802B is remarkably close to the brown
dwarf/substellar boundary, and more measurements with this technique
will provide tight constraints on the substellar evolutionary models. 

\acknowledgements

We thank Rick Burruss, Jeff Hickey and the Palomar observatory staff
for help with these observations.  This work is partially funded by
the National Science Foundation under grants AST-0335695 and
AST-0506588.
The research described in this paper was performed in part by the Jet
Propulsion Laboratory, California Institute of Technology, under contract with
the National Aeronautics and Space Administration.
This publication makes use of data products from the Two Micron All
Sky Survey, which is a joint project of the University of Massachusetts
and the Infrared Processing and Analysis Center/California Institute of
Technology, funded by the National Aeronautics and Space Administration
and the National Science Foundation.
 
\bibliographystyle{apj}
\bibliography{gj802}

\end{document}